\documentclass[aps,prb,reprint,superscriptaddress,longbibliography]{revtex4-2}

\usepackage{amsmath,amssymb,bm}
\usepackage{graphicx}
\usepackage[dvipdfmx]{hyperref}
\usepackage{physics}
\usepackage{color}
\usepackage{comment}

\begin{document}

\title{Topological Filtering and Emergent Kondo Scale}

\author{Ryosuke Yoshii}
\affiliation{Center for Liberal Arts and Sciences, Sanyo-Onoda City University, Onoda Yamaguchi 756-0884, Japan}
\email{ryoshii@rs.socu.ac.jp}

\author{Rio Oto}
\affiliation{Department of Engineering, Sanyo-Onoda City University, Onoda, Yamaguchi 756-0884, Japan}

\date{\today}

\begin{abstract}
We study the Kondo effect induced by a topological soliton in a one-dimensional Dirac system with the sign-changing mass term. 
The soliton hosts a localized zero mode whose spatially extended wavefunction leads to a momentum-dependent exchange coupling with itinerant electrons. 
We show that this structure generates a nontrivial form factor that suppresses high-energy scattering processes, resulting in an energy-dependent effective Kondo coupling. 
As a consequence, the real-space structure of the soliton directly controls the emergent Kondo scale.
This work establishes a mechanism by which topological defects control many-body energy scales through their wavefunction structure, suggesting a general principle for engineering many-body energy scales via topology. 
\end{abstract}

\maketitle

\section{Introduction}

Topological defects are known to host localized electronic states protected by band topology, as exemplified by soliton solutions in one-dimensional systems \cite{JackiwRebbi,SSH,Takayama,Campbell1981,Campbell1982,Heeger}.
While such states are well understood at the single-particle level, their role in controlling emergent many-body energy scales remains largely unexplored. 
A central open question is whether the real-space structure of a topological bound state can directly determine the scale of many-body correlations. 

The Kondo effect provides a canonical example of an emergent low-energy scale generated by logarithmic renormalization of impurity scattering \cite{Kondo,Anderson,Wilson,Hewson}. 
In conventional settings, the impurity is effectively point-like, leading to a momentum-independent coupling and a Kondo temperature set by the electronic bandwidth. 
Extensions to systems with an energy-dependent density of states have revealed qualitatively new scaling behavior \cite{WithoffFradkin,GonzalezBuxton}. 
Such situations arise, for example, in graphene and topological edge systems, where the vanishing or structured density of states leads to unconventional Kondo physics \cite{Uchoa,Maciejko}. 
In all these cases, however, the energy dependence originates from properties of the bath. 

In this work, we demonstrate a distinct and generic mechanism in which the energy dependence arises instead from the impurity wavefunction itself. 
We show that a spatially extended bound state acts as a momentum-space filter that suppresses high-energy scattering processes, thereby defining its own effective ultraviolet scale. 
Importantly, the form factor is not an arbitrary property of an extended impurity, but is fixed by the topology of the underlying Dirac mass domain wall. 
The algebraic decay $\mathcal{F}(\epsilon)\sim \epsilon^{-4}$ originates from the Fourier transform of the exponentially localized zero mode in one dimension. 
While the precise power depends on the spatial profile of the bound state, the existence of a rapidly decaying form factor is a generic consequence of topological localization. 

As a concrete realization, we consider a one-dimensional Dirac system with a sign-changing mass, which hosts a topological soliton with a localized zero mode (see Fig.~\ref{fig:schematic}). 
The Fourier transform of this mode generates a form factor that suppresses scattering processes with energies $\epsilon \gg m$, where $m$ is the soliton mass. 
As a result, the logarithmic renormalization is effectively truncated at $\epsilon \sim m$, replacing the bare bandwidth by an emergent, topology-controlled ultraviolet scale. 

This leads to a Kondo temperature of the form
\begin{equation}
T_K \sim m \exp(-A m^2),
\end{equation}
where $A \sim g/(64\rho V^2)$. 
Here, the topological mass controls both the prefactor and the exponent of the many-body scale: it enhances interactions through spatial confinement while suppressing hybridization via wavefunction delocalization. 
Our results establish a general mechanism by which the spatial structure of topological bound states governs emergent many-body energy scales. 

Importantly, the present mechanism is fundamentally different from both pseudogap Kondo systems and conventional extended impurities. 
In pseudogap systems, the energy dependence originates from the bath density of states, whereas in our case it arises solely from the impurity wavefunction. 
Moreover, unlike generic extended impurities where the form factor is non-universal, here it is uniquely fixed by topology.

\section{Model and Solitonic Zero Mode}

\begin{figure}
\centering
\includegraphics[clip,width=8.8cm]{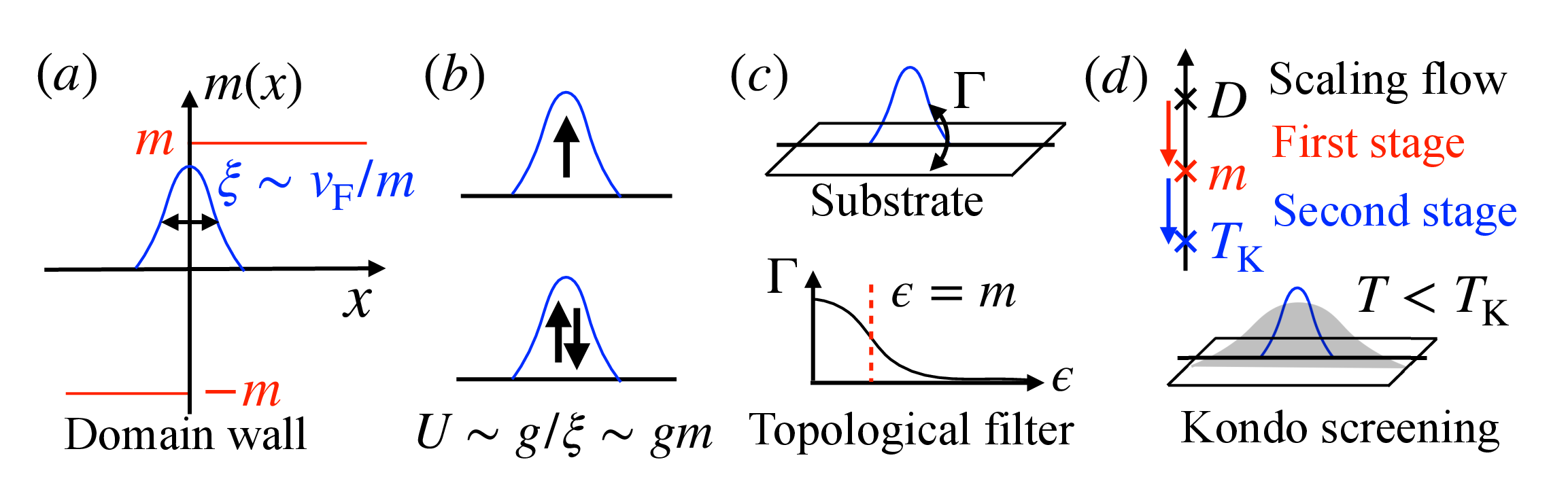}
\caption{Schematic of the solitonic Kondo system. 
The  Dirac system with a mass that changes the sign at $x=0$ from $-m$ to $m$ is considered (Fig.~(a)). 
At the position where the mass changes sign, a solitonic mode is localized with the localization length $\xi \sim 1/m$. 
 In the localized mode, the strength of the Coulomb interaction $U$ becomes $U\propto gm$, which inhibits the double occupation in the solitonic mode and results in the formation of a local moment (Fig.~(b)). 
If this system is attached to another bulk system, such as the substrate, the hybridization takes place and its strength $\Gamma$ highly depends on the shape of the soliton (Fig.~(b)). 
This hybridization works as a low-pass filter and is thus called a topological filter. 
Due to the topological filter, the high energy region down from bandwidth $D$ to $m$ is cutoff and the relevant UV scale for the Kondo effect becomes $m$, resulting in a significant $m$ dependence of the Kondo temperature $T_{\rm K}$ (Fig.~(d)). 
}
\label{fig:schematic}
\end{figure}

We consider a one-dimensional Dirac Hamiltonian with the Fermi velocity $v_{\rm F}$
\begin{equation}
H_D = \int dx\, \Psi^\dagger(x)
\left[- i \hbar v_{\rm F} \partial_x \sigma_z + m(x)\sigma_x \right]
\Psi(x),
\end{equation}
where $m(x)$ changes sign across a domain wall,
\begin{equation}
m(x) = m\, \mathrm{sgn}(x).
\end{equation}
For simplicity, we set $\hbar=1$ and $v_{\rm F}=1$. 
This configuration hosts a Jackiw--Rebbi solitonic zero mode localized at the defect as schematically shown in Fig.~\ref{fig:schematic} (a).
The shape of the soliton becomes (See Appendix \ref{derivation} for details)
\begin{equation}
\phi_0(x) =\sqrt{m}\, e^{-m |x|}.
\end{equation}
The soliton localization length is thus $\xi = 1/m$.

We decompose the fermionic field as
\begin{equation}
\hat\Psi_\sigma (x) :=  \hat \phi_0(x)\hat d_\sigma + \sum_k \hat \phi_k(x) \hat c_{k,\sigma},
\end{equation}
where $\hat d_\sigma$ annihilates the solitonic zero mode with spin $\sigma$ and $\hat c_{k,\sigma}$ describe extended states with wave number $k$ and spin $\sigma$. 
We also define $\hat\Psi(x):=\sum_\sigma \hat\Psi_\sigma (x)$. 

While the present discussion is based on the continuum Dirac model, the essential structure of the effective impurity model is not restricted to this limit. 
A similar construction starting from a lattice model, such as the Su-Schrieffer-Heeger (SSH) chain supporting topological solitons, leads to an equivalent Anderson impurity model, demonstrating that the mechanism is not tied to the continuum Dirac description. 
For completeness, we summarize this derivation in Appendix \ref{derivation}.

\begin{figure}[t]
\centering
\includegraphics[clip,width=8.5cm]{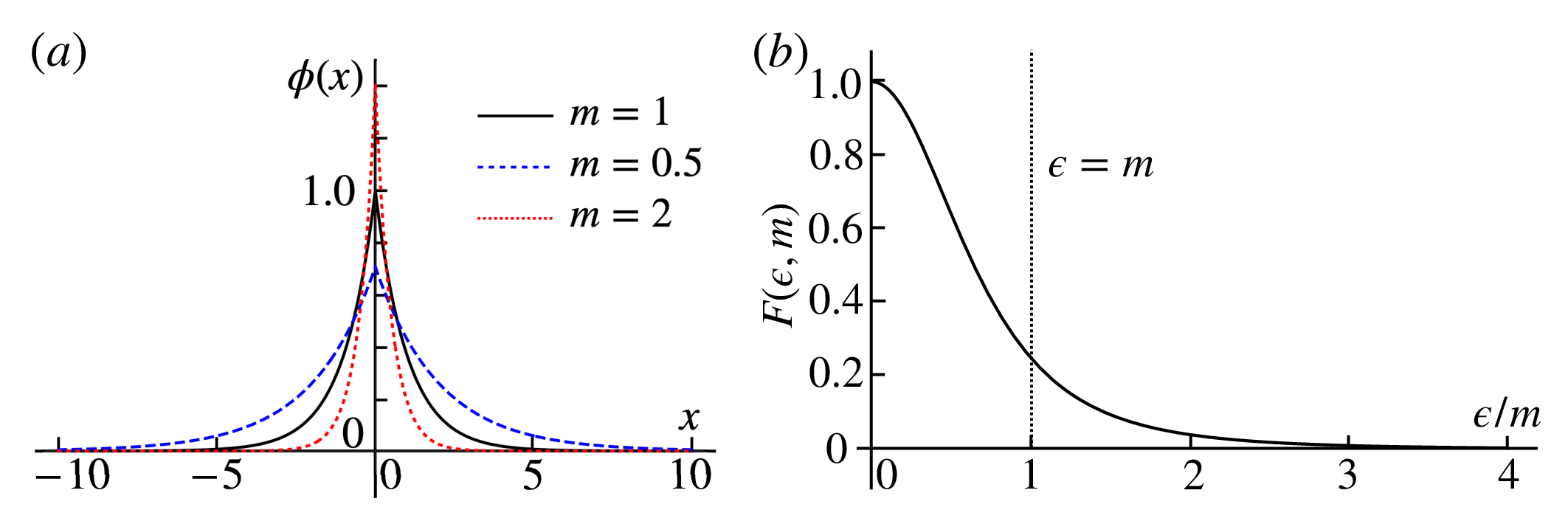}
\caption{
Solitonic mode as a function of $x$ (Fig.(a)) and topological filter emerged from the solitonic mode (Fig.(b)). 
The solitonic mode is localized in the vicinity of the domain wall placed at $x=0$. 
The topological mass controls the localization length as depicted in Fig.(a) (solid, dashed, and dotted lines correspond to $m=1$, $m=0.5$, and $m=2$, respectively). 
In Fig.~(b), the topological filter $F(\epsilon, m)$ defined in Eq.~\eqref{topfilter} is plotted as a function of $\epsilon/m$. The rapid $\epsilon^{-4}$ decay for $\epsilon \gg m$ demonstrates that the high-energy itinerant states are effectively decoupled from the soliton spin, allowing the topological mass $m$ to function as a physical UV cutoff for the Kondo scaling.
}
\label{topofilter}
\end{figure}

\section{Emergent Anderson Impurity}

Including a local interaction
\begin{equation}
H_{\rm int} = \frac{g}{2} \int dx\, 
\left(\hat \Psi^\dagger \hat \Psi\right)^2 ,
\end{equation}
and projecting onto the zero mode yields an effective Coulomb interaction
\begin{equation}
U_{\rm eff} = g \int dx\, |\phi_0(x)|^4
= \frac{g m}{2}.
\end{equation}
The soliton, therefore generates a localized correlated state (Fig.~\ref{fig:schematic} (b)).
The interpretation of this form is quite simple: the electrons confined in the solitonic mode feel the Coulomb repulsion, which is proportional to the inverse of the confinement length $\xi \sim 1/m$.

We now couple the chain to a metallic substrate,
$ H_{\rm sub} = \sum_{k\sigma} \epsilon_k
\hat a^\dagger_{k\sigma} \hat a_{k\sigma}$, 
via tunneling
\begin{equation}
H_{\rm hyb} = V\sum_\sigma\int dx\,
\hat \Psi_\sigma^\dagger(x) \hat \chi_\sigma(x) + \mathrm{h.c.},
\end{equation}
where $\hat \chi_\sigma(x)=\sum_k \hat a_{k\sigma}e^{ikx}$ denotes substrate electrons. 
Such coupling naturally arises when the Dirac system
is placed on a metallic substrate or contacted by
metallic electrodes. 
Here, the metallic bath is required to provide low-energy screening electrons for the emergent impurity moment. 
Projecting onto the zero mode gives
\begin{equation}
H_{\rm hyb}^{(0)} = 
\sum_{k\sigma} V_k(m)\,
\hat a^\dagger_{k\sigma} \hat d_\sigma + \mathrm{h.c.},
\end{equation}
with
\begin{equation}
V_k(m) \propto V
\int dx\, \phi_0(x) e^{ikx}
= \frac{2V}{\sqrt{m}} \frac{m^2}{m^2+k^2}.
\label{Vkm}
\end{equation}

By recasting the relevant terms, we arrive at 
\begin{align}
H=&\sum_{\sigma} \epsilon_d \hat d^\dagger_{\sigma}\hat d_{\sigma}+ U_{\rm eff} \hat n_{\uparrow}\hat n_{\downarrow}
+ \sum_{k\sigma} \epsilon_k \hat a^\dagger_{k\sigma} \hat a_{k\sigma}\nonumber\\
&+\sum_{k\sigma} V_k \hat a^\dagger_{k\sigma} \hat d_\sigma + \mathrm{h.c.},
\label{eq:AndersonModel}
\end{align}
where $\epsilon_d$ is the energy difference between the Fermi level of the substrate and the center of the gap in the system. 
The soliton zero mode forms an effective spin-$1/2$ degree of freedom when the on-site interaction $U_{\rm eff}$ is sufficiently large, suppressing charge fluctuations and stabilizing single occupancy. 
The hybridization $\Gamma(\epsilon)$ between the solitonic mode and the substrate is given as 
\begin{equation}
\Gamma(\epsilon)=\pi \rho |V_k|^2=\frac{4\pi \rho V^2 m^3}{(m^2+\epsilon^2)^2}. 
\label{Gammaepsilon}
\end{equation}
This follows from the Fourier transform of the exponential bound-state wavefunction. 
This energy-dependent hybridization also has a simple physical interpretation. 
The high-energy mode, or equivalently the short-wavelength mode, cannot feel the solitonic mode, and thus the hybridization is negligibly small for $\epsilon \gg m$. 
On the other hand, the solitonic mode becomes a point-like defect for the low-energy (long-wavelength) mode, which has the hybridization whose strength is proportional to the localization length $\xi=1/m$ for $\epsilon \ll m$. 

We make a few remarks on this effective model. 
For this effective model, we can make a criterion to get a localized moment. 
Since the energy of the doubly occupied state must be much larger than the level-width of the energy level, $U_{\rm eff}/\Gamma(0)\gg 1$ must be satisfied. 
This condition is satisfied when $m$ is large compared to the inverse lattice scale. 
We also note that the Coulomb interaction is defined in the continuum limit, but it should depend on the lattice constant $a$ as $g\sim U_0 a$, and the effective interaction is given by the ratio of the lattice constant and the confinement length $\xi$ as $U_{\rm eff}\sim U_0 am=U_0a/\xi  $. 
We emphasize that the emergence of an effective Anderson impurity model is not specific to the continuum Dirac description.
An analogous derivation starting from a discrete SSH chain hosting a topological soliton leads to the same structure of the impurity Hamiltonian, as summarized in Appendix \ref{discretemodel}.

\section{Topological filter and Emergent UV Scale}

The structure of the hybridization function $\Gamma(\epsilon) \propto (m^2+\epsilon^2)^{-2}$ decays rapidly as $\epsilon^{-4}$ for $\epsilon \gg m$. 
The $\epsilon^{-4}$ decay originates from the square of the Fourier transform of the exponentially localized soliton wavefunction, and is therefore a direct consequence of the topological bound state.
This motivates the definition of the topological filter $\mathcal{F}(\epsilon, m)$ as
\begin{equation}
\mathcal{F}(\epsilon, m) := \left[1+(\epsilon/m)^2\right]^{-2},
\label{topfilter}
\end{equation}
where the hybridization is expressed as $\Gamma(\epsilon)=\Gamma(0) \mathcal{F}(\epsilon, m)$. 
As seen from Eq.~\eqref{Vkm}, this filter originates from the form factor of the solitonic state. 
Due to this rapid decay, high-energy processes with $\epsilon \gg m$ are suppressed, and the renormalization flow effectively starts from the scale $m$. 
In contrast to the conventional Anderson model with constant hybridization, the rapid $\epsilon^{-4}$ decay ensures that contributions from $\epsilon \gg m$ are power-law suppressed, so that the logarithmic divergence effectively truncates at $\epsilon \sim m$ instead of the bare bandwidth $D$. 
This structure naturally leads to an effective two-stage scaling picture discussed in the following section. 
Unlike generic extended impurities, the existence and qualitative structure of the form factor here are dictated by topology, as they originate from the topological zero mode associated with the bulk mass domain wall. 

In Fig.~\ref{topofilter}, we plot the spatial configuration of the solitonic mode for various $m$ (Fig.~\ref{topofilter}(a)) and $\mathcal{F}(\epsilon,m)$ as a function of $\epsilon/m$ (Fig.~\ref{topofilter}(b)). 
This result shows that the topological filter decouples the high-energy itinerant states from the solitonic state.

\section{Two-stage Scaling Analysis} 

The renormalization group (RG) flow of the present system is uniquely characterized by the topological mass $m$. 
While the conventional Kondo flow starts from the bandwidth $D$, our topological filter $\mathcal{F}$ effectively dictates a two-stage evolution: the high-energy flow ($D>\epsilon>m$) is governed by the m-dependent suppression of scattering, which then provides a topologically tuned initial condition for the subsequent Kondo scaling ($ \epsilon < m$). 

The crossover scale $m$ appearing in the two-stage scaling procedure has a simple physical interpretation. 
The soliton wavefunction has a spatial extent $\xi\sim 1/m$, which suppresses the coupling to conduction electrons with wavelengths shorter than the soliton size. 
Consequently, electronic states with energies larger than $m$ contribute weakly to the hybridization,
and the renormalization flow effectively starts at the scale $m$. 

To elucidate how topology controls the emergent many-body scale, we perform a two-stage scaling analysis based on the effective Anderson model derived in Eq.~\eqref{eq:AndersonModel} \cite{Haldane}. 
The central feature of our model is the strong energy dependence of the hybridization $V_{k}$, which originates from the spatial profile of the topological soliton. 
For more detailed analysis on the scaling equation, see Appendix \ref{RGevaluation}. 

Physically, the first stage eliminates virtual processes that probe length scales shorter than the soliton size, while the second stage describes spin fluctuations that fully resolve the soliton as a localized moment.

\subsection{Stage 1: Renormalization of the impurity level}
In the first stage, we integrate out high-energy charge fluctuations from the microscopic bandwidth $D$ down to the topological mass scale $m$. 
The impurity level is renormalized by virtual charge fluctuations, leading to
\begin{equation}
\delta \epsilon_d = \int d\epsilon \frac{\Gamma(\epsilon)}{\epsilon_d - \epsilon}.
\end{equation}
In contrast to the conventional case with constant hybridization, the soliton-induced form factor suppresses high-energy contributions as $\Gamma(\epsilon) \sim \epsilon^{-4}$. 
As a result, the integral is convergent and dominated by energies $\epsilon \lesssim m$, yielding a finite renormalization of order $\ln m$. 
More precisely, the logarithmic renormalization integral $\int d\epsilon\, \Gamma(\epsilon)/\epsilon$  saturates at $\epsilon \sim m$, 
so that contributions from $\epsilon \gg m$ become negligible.
The renormalized impurity level sets the effective exchange coupling via the Schrieffer-Wolff transformation, linking charge fluctuations to the Kondo scale. 
This shows that charge fluctuations are effectively confined to the same energy window that governs the Kondo scaling, reinforcing the role of $m$ as the emergent ultraviolet scale.
This stage effectively prepares the initial condition for the Kondo scaling at the scale $m$. 
Physically, this stage corresponds to integrating out charge fluctuations that are spatially incompatible with the extended soliton wavefunction. 

\subsection{Stage 2: Poor man's scaling and Kondo temperature}
In the second stage, we map the system onto an effective Kondo model via the Schrieffer-Wolff transformation. The effective exchange coupling $J(m)$ at the scale $m$ is determined by the hybridization at the Fermi level and the effective interaction $U_{\rm {eff}} = gm/2$:
\begin{equation}
    \rho J(m) \approx \frac{8 \rho |V_k(0)|^2}{U_{\rm {eff}}} = \frac{64 \rho V^2}{g m^2},  
    \label{Kondocoupling}
\end{equation}
where we have assumed particle-hole symmetry for simplicity. 
This expression shows that both the ultraviolet cutoff and the effective coupling are controlled by the same topological parameter m. 
The strong $m^{-2}$ dependence originates from two independent effects.
The effective interaction scales as $U_{\rm eff} \propto m$ due to the localization of the soliton wavefunction,
while the effective hybridization $|V_{\rm eff}(0)|^2 \propto 1/m$ reflects the spatial extent $\xi \sim 1/m$ of the soliton state.
Together these contributions yield $ J(m) \propto m^{-2}$. 
These scalings originate from distinct physical mechanisms: interaction enhancement due to confinement and hybridization suppression due to spatial extension. 
Evaluating the one-loop vertex correction while retaining the momentum dependence of the exchange coupling, we obtain the scaling equation for the dimensionless coupling $g_{\rm {eff}}(\epsilon) = \rho J(\epsilon)$ as
\begin{equation}
    \frac{d g_{\text{eff}}(\epsilon)}{d \ln \epsilon} = - g_{\rm {eff}}(\epsilon)^2 \mathcal{F}(\epsilon, m).
    \label{poormanscaling}
\end{equation}
This scaling equation is derived from the one-loop vertex correction while explicitly retaining the momentum dependence of the exchange coupling, followed by an energy-shell average over the running cutoff (see Appendix \ref{RGevaluation}). 
Here, the momentum dependence of the exchange coupling is incorporated through an energy-shell average, yielding an effective flow equation controlled by the squared form factor.
The topological filter $\mathcal{F}(\epsilon,m)$ modifies the phase space of these processes. 
As shown in Appendix \ref{RGevaluation}, the logarithmic integral explicitly yields $\ln(m/T_K)$, confirming that $m$ replaces the bare bandwidth as the ultraviolet scale and thus $T_{\rm K}$ becomes 
\begin{equation}
    T_{\rm K}(m) \approx \mathcal{C} m \exp \left( - \frac{1}{\rho J} \right) = \mathcal{C}m \exp \left( - \frac{g m^2}{64 \rho V^2} \right),
    \label{TKapproximated}
\end{equation}
where $\mathcal{C}$ is a dimensionless constant. 
The same scaling form is obtained for a parabolic dispersion, confirming that the exponential dependence of $T_{\rm K}$ is not tied to the linear dispersion assumed in the main text (see Appendix \ref{parabolic}).
The scaling can be understood as
\begin{equation}
T_K \sim (\text{topological UV scale}) \times \exp(-1/\rho J(m)),
\end{equation}
where both factors are controlled by the same topological parameter.

Importantly, the same scaling structure is reproduced within the s-d model, demonstrating that the emergence of the topological UV scale is not tied to the Anderson representation but is a robust consequence of the soliton-induced form factor (see Appendix \ref{sdmodel}). 
The spatial profile of the soliton wavefunction induces a momentum-dependent exchange coupling $J_{k,k^\prime}=Jf(q)$ with  $f(q)=\frac{m^2}{m^2+q^2}$, 
where $q=(k-k^\prime)/2$. 
This form factor suppresses high-energy scattering processes and effectively introduces an ultraviolet cutoff set by the soliton mass $m$. 
The scaling equation for $J$ becomes $dJ/d\ln \epsilon=-\rho J^2 f(q)^2$ within a one-loop approximation, which is consistent with Eq.~\eqref{poormanscaling}.  
This scaling equation is derived from the one-loop vertex correction while explicitly retaining the momentum dependence of the exchange coupling. 
Averaging over states within the running cutoff yields an effective flow equation governed by the squared form factor.

\begin{figure}
\centering
\includegraphics[clip,width=8.6cm]{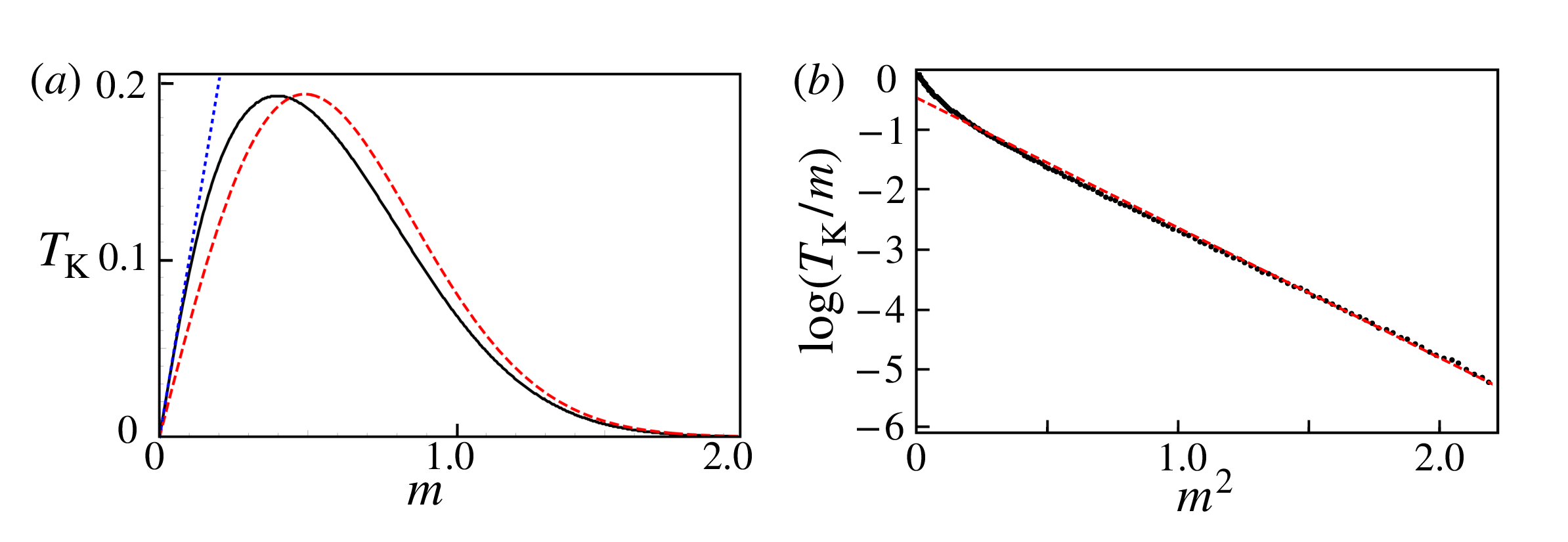}
\caption{Topological control of the Kondo temperature. 
Kondo temperature $T_{\rm K}$ is plotted as a function of the topological mass $m$ for $g=2.0$ and $\rho V^2 = 0.15$ (Fig.~(a)). 
The numerical results obtained by scaling flow that starts from the initial coupling $g_0=\rho J(m)$ (black solid line) are in excellent agreement with the analytical formula $T_K \approx \mathcal{C} m \exp(-Am^2)$ (red dashed line) derived in Eq.~\eqref{TKapproximated}. 
The prefactor $\mathcal{C}$ is expected to depend on subleading corrections beyond the one-loop scaling analysis and is therefore treated as a non-universal constant. 
The characteristic dome-shaped behavior emerges from the competition between the localization-induced enhancement of the effective interaction $U_{\text{eff}} \propto m$ and the suppression of the hybridization volume. 
The blue dotted line $T_{\rm K} = m$ indicates the theoretical upper bound; beyond this limit (where $T_{\rm K} \approx m$), the system is expected to crossover into a valence-fluctuation regime. 
Figure (b) shows $\log(T_{\rm K}/m)$ as a function of $m^2$ (solid line). 
The dashed line is the linear fit of the data (here the slope is $\sim -2.193$).   
}
\label{fig:TKVSm}
\end{figure}

\subsection{Physical implications}
The result in Eq.~\eqref{TKapproximated} reveals profound effects of topology on the Kondo scale. 
First, the prefactor of $T_{\rm K}$ is determined by the topological mass $m$ instead of $D$, indicating that topology generates its own UV scale. 
This ``topological filtering" of high-energy modes makes the Kondo effect insensitive to microscopic details of the lead. 
Second, $T_{\rm K}$ exhibits an extreme sensitivity to $m$ due to the $m^2$ dependence in the exponent. 
This sensitivity provides a natural explanation for the large heterogeneity of $T_{\rm K}$ observed in experiments~\cite{Li2020, Zhao2005}, as small local variations in the topological gap $m$ are exponentially amplified in the many-body scale.
Third, there is an optimal value of $m$ at which the Kondo temperature is maximized.

\section{Numerical Results and Validity Regime}

To verify the analytical scaling law derived in Eq.~(\ref{TKapproximated}), we perform a numerical integration of the scaling equation. 
The purpose is to confirm the universality of the scaling structure, rather than to obtain quantitatively precise values of the Kondo temperature.
We set the hybridization strength to $\rho V^2 = 0.15$, which corresponds to a realistic value of $\Gamma \approx 47$~meV for a bandwidth of $D = 1$~eV, placing the system in a regime accessible to experimental observation in topological insulators or graphene nanoribbons.

Figure \ref{fig:TKVSm} shows the Kondo temperature $T_{\rm K}$ as a function of the topological mass $m$. The numerical results (black solid line) clearly exhibit a non-monotonic dome-like behavior, confirming the competition between the topological prefactor and the exponential suppression due to soliton localization.
The analytical result based on Eq.~(\ref{TKapproximated}) (red dashed line) is also plotted. 
The theoretical upper bound, beyond which (where $T_{\rm K} \approx m$), the system is expected to crossover into a valence-fluctuation regime, is indicated by the dotted line ($T_{\rm K} = m$). 
The agreement is obtained without introducing additional fitting parameters beyond an overall constant prefactor. 

In the region $T_{\rm K} \ll m$, the separation of scales is well-maintained, and the topological filter $\mathcal{F}(\epsilon, m)$ effectively redefines the renormalization process. 
As $m$ increases beyond the optimal value $m_{\rm opt} \approx \sqrt{32 \rho V^2 / g}$, the rapid increase of the effective local repulsion $U_{\rm eff} = gm/2$ and decrease of hybridization $\Gamma\propto 1/m$ lead to the exponential decay of $T_{\rm K}$. 

Crucially, our numerical results show a slight shift of the peak position compared to the leading-order analytical prediction $m_{\rm opt} \approx \sqrt{32 \rho V^2 / g}$. 
This discrepancy arises from the continuous nature of the topological filter, which is naturally captured by the numerical integration but neglected in the simple analytical ansatz. 
This robustness of the ``topological optimization'' across both analytical and numerical methods underscores the feasibility of controlling many-body scales via band-structure topology. 
The precise prefactor of the effective exchange coupling depends on the microscopic structure of the hybridization matrix elements, but the scaling $J(m)\propto m^{-2}$ captures the leading dependence on the soliton mass.
The numerical solution confirms that the analytical scaling form remains valid even beyond the regime where the sharp cutoff approximation breaks down.

We note that the Kondo fixed point itself is identical to the conventional Anderson impurity problem.
The topological defect modifies only the ultraviolet structure of the hybridization function, while the low-energy strong-coupling physics remains universal.
Therefore, the present mechanism can be captured within the scaling framework without requiring a full numerical renormalization-group calculation. 
A full numerical renormalization-group calculation would provide a quantitatively precise determination of the prefactor, but is not expected to modify the scaling structure derived here. 

\section{Conceptual Implications}

Equation \eqref{TKapproximated} demonstrates that topology controls the entire many-body scale.
In conventional Kondo systems, the ultraviolet cutoff and exchange coupling
are externally defined microscopic parameters.
Here, both the prefactor and the exponent are governed by the topological mass.

The dependence $T_{\rm K} \sim \mathcal{C} m \exp(-Am^2)$ is found. 
In the limit $m \to 0$, the soliton delocalizes and the Kondo scale
vanishes linearly in $m$, whereas $m\to \infty$ again gives exponential decay towards $T_{\rm K}\to 0$. 
Topology, therefore, does not merely create a localized state; it controls the exponential generation of correlations.

This mechanism differs fundamentally from pseudogap Kondo physics,
where the bath density of states determines the critical behavior.
In the present case, the metallic substrate provides a finite density of states,
while the defect topology controls the impurity coupling itself.
The soliton thus transmutes band topology into a nonperturbative correlation scale. 
This interpretation is further supported by considering the Wilson chain representation of the impurity problem. 
The energy-dependent hybridization leads to rapidly decaying hopping amplitudes along the chain, effectively truncating the chain at the scale set by the soliton mass. 
This provides a non-perturbative justification for identifying m as the emergent ultraviolet scale.

Our results indicate that topological defects can act as generators of correlated impurity physics. The resulting Kondo scale is determined by the localization scale of the defect-bound state, suggesting a general link between topology and emergent many-body energy scales. 
This establishes a new paradigm where topology determines not only the existence of localized states but also the scale of many-body correlations. 

\section{Experimental Realization and Tunability of $m$}

The predicted scaling of the Kondo temperature provides a direct route to experimentally test the role of the impurity wavefunction in renormalization. 
The essential requirement is a system hosting a localized bound state whose spatial extent can be tuned in a controlled manner. 
Among possible platforms, graphene nanoribbons and topological crystalline insulator surfaces are particularly promising systems in which both solitonic bound states and Kondo signatures have been experimentally observed.

A natural realization is provided by Dirac systems with a spatially varying mass, where localized bound states emerge. Such systems have been realized experimentally, and their local electronic structure can be probed using scanning tunneling spectroscopy \cite{ExpDiracSTM}.

The Kondo temperature can be extracted from the width of the zero-bias resonance in tunneling measurements \cite{ExpKondoSTM}. By systematically varying the mass scale $m$, for instance, through electrostatic gating, magnetic proximity effects, or controlled domain-wall engineering, one can directly test the predicted scaling relation 
$ T_K \sim m \exp(-A m^2)$.

A particularly clear signature is obtained by plotting $\ln(T_K/m)$ as a function of $m^2$, which is expected to exhibit a linear dependence. This behavior is qualitatively distinct from conventional Kondo systems, where the bandwidth sets the ultraviolet scale and no dependence on a tunable spatial parameter appears.

Observation of this scaling would provide direct evidence that the real-space structure of the impurity wavefunction governs the renormalization flow.

\section{Conclusion}

We have shown that a solitonic topological defect embedded in a metal realizes a defect-generated impurity whose many-body scale is dictated by topology.
The topological soliton acts as a momentum-space form factor that suppresses high-energy scattering. 
Importantly, the soliton mass does not act as a sharp cutoff but as an effective ultraviolet scale: contributions from energies much larger than $m$ are strongly suppressed by the form factor, rendering the renormalization insensitive to the microscopic bandwidth. 
The resulting Kondo temperature obeys
$T_{\rm K}(m) = \mathcal{C} m \exp(-Am^2)$,
demonstrating exponential control of correlations by the topological mass.
This establishes a general paradigm in which topological defects govern emergent many-body scales in a fundamentally nonperturbative manner.

Conceptually, the present mechanism shows that topology can generate its own ultraviolet scale for many-body renormalization.
In contrast to conventional Kondo systems, where the bandwidth sets the cutoff, here the topological mass determines the energy window in which impurity screening develops. 
This mechanism suggests a general route to engineer many-body energy scales through the spatial structure of topological bound states. 
The present mechanism is expected to apply broadly to impurity states with finite spatial extent, beyond the specific Dirac realization considered here. 
This generality is further supported by the fact that an equivalent mechanism arises in lattice realizations such as the SSH chain, where the solitonic bound state leads to the same form-factor-controlled hybridization structure.

\section*{acknowledgement}
This work is supported by JSPS KAKENHI No.~JP25K07156.

\appendix

\section{Dirac equation in one-dimensional system with spatial-dependent mass}
\label{derivation}

We consider a one-dimensional interacting model with a spatially dependent mass that changes the sign described by the Hamiltonian  
\begin{align}
H=&\int dx \left\{\sum_\sigma \hat \psi_\sigma^\dagger(x) {\mathcal H}_{\rm D} \hat \psi_\sigma(x)+\frac{g}{2} \left[\sum_\sigma \hat \psi_\sigma^\dagger(x) \hat \psi_\sigma(x)\right]^2\right\}, 
\label{eq:model}
\end{align}
where the Hamiltonian density ${\mathcal H}_{\rm D}$ is defined as 
\begin{align}
&{\mathcal H}_{\rm D}:=-i\partial_x \sigma_z+m(x)\sigma_x, 
\label{eq:Dirac}\\
&m(x):=m\cdot {\rm sgn}(x),
\label{msign}
\end{align}
where $\sigma_x$ and $\sigma_z$ are the $x$ and $z$ component of the Pauli matrices, respectively. 
The two component spinor $\psi_\sigma$ ($\sigma=\uparrow$, $\downarrow$) is defined by 
\begin{align}
\hat \psi_{\sigma}(x)=\left(
\begin{array}{cc}
\hat \psi_{A,\sigma}(x) \\
\hat \psi_{B,\sigma}(x)
\end{array}
\right),  
\label{eq:spinor}
\end{align}
where $\hat\psi_{A/B, \sigma } (x)$ represents the annihilation operator for the electron with spin $\sigma$ in sublattice $A/B$ at position $x$. 

First we diagonalize the first term in Eq.~\eqref{eq:model} by solving the following eigenvalue equation 
\begin{align}
&{\mathcal H}_{\rm D}\hat \psi_\sigma(x)=E\hat\psi_\sigma(x),
\label{eq:eigeneq}
\end{align}
Let us find the zero mode solution, namely the solution of \eqref{eq:eigeneq} with $E=0$.  
Multiplying $\sigma_z$ from the left, Eq.~\eqref{eq:eigeneq} for $E=0$ becomes  
\begin{align}
\partial_x \hat\psi_{\sigma}(x)=m(x)\sigma_y \hat\psi_{\sigma}(x).  
\label{eq:zeromode}
\end{align}
This equation has a solution with the following form 
\begin{align}
\hat \psi_{\sigma}(x)=\frac{1}{\sqrt{2}}\left(
\begin{array}{cc}
1 \\
\pm i
\end{array}
\right)\phi_{\pm}(x)\hat d_{\pm,\sigma}, 
\label{eq:zeromodesol}
\end{align}
where $\phi_\pm(x)$ and $\hat d_{\pm,\sigma}$ are the zero-mode function and the annihilation operator for that with spin $\sigma$, respectively. 
Substituting Eq.~\eqref{eq:zeromodesol} into Eq.~\eqref{eq:zeromode}, we find 
\begin{align}
\partial_x  \phi_{\pm}(x)=\pm m(x) \phi_{\pm}(x).  
\label{eq:eqphipm}
\end{align}
The solution of Eq.~\eqref{eq:eqphipm} becomes 
\begin{align}
\phi_{\pm}(x)=A_{\pm}e^{\pm \int^x m(y)dy}=A_{\pm}e^{\pm m |x|},   
\label{eq:zeromodespm}
\end{align}
where $A_{\pm}$ is the normalization factor. 
Though we find two solutions, $\phi_{+,\sigma}(x)$ diverges in the infinite system. 
Thus, we obtain the unique zero-mode solution localized in the vicinity of $x=0$ as 
\begin{align}
\phi_{-,\sigma}(x)=\sqrt{m}e^{- m |x|}.     
\label{eq:zeromode}
\end{align}
Finally, we obtain the zero-mode solution $\hat \psi_{0,\sigma}(x)$ as 
\begin{align}
\hat \psi_{0,\sigma}(x)=\sqrt{\frac{m}{2}}\left(
\begin{array}{cc}
1 \\
-i 
\end{array}
\right)e^{- m |x|} \hat d_\sigma, 
\label{eq:zeromodesol2}
\end{align}
where we drop the subscription $-$ of $\hat d_\sigma$ for simplicity. 

By separating the zero mode, we can expand $\hat \psi_{\sigma}(x)$ as 
\begin{equation}
\hat \psi_{\sigma}(x)=\hat \psi_{0,\sigma}(x)+\sum_{k\neq 0} \hat \psi_{k,\sigma}(x). 
\label{eq:modeexpansion}
\end{equation}
The substitution of Eq.~\eqref{eq:modeexpansion} into the second term in Eq.~\eqref{eq:model} yields 
\begin{align}
&\frac{g}{2}\int dx \left\{ \left[\sum_\sigma \hat \psi_{0,\sigma}^\dagger(x) \hat \psi_{0,\sigma}(x)\right]^2 
+   \sum_k\left[\sum_\sigma \hat \psi_{0,\sigma}^\dagger(x) \hat \psi_{k,\sigma}(x)\right]^2 \right.\nonumber\\
&\left.+   \sum_k\left[\sum_\sigma\hat \psi_{k,\sigma}^\dagger(x) \hat \psi_{0,\sigma}(x)\right]^2 + \sum_{k,k^\prime}\left[\sum_\sigma\hat \psi_{k,\sigma}^\dagger(x) \hat \psi_{k^\prime,\sigma}(x)\right]^2 \right\}. 
\label{eq:interactionterm}
\end{align}
The first term of Eq.~\eqref{eq:interactionterm} becomes  
\begin{align}
g\int dx  m^2 e^{-4m |x|} \hat d_{\uparrow}^\dagger \hat d_{\uparrow} \hat d_{\downarrow}^\dagger \hat d_{\downarrow}  =\frac{gm}{2} \hat d_{\uparrow}^\dagger \hat d_{\uparrow} \hat d_{\downarrow}^\dagger \hat d_{\downarrow}. 
\end{align}
Since the normalization factors of the second and third terms are $m/L$, the integration yields $\sim 1/L$ for the system size $L$, and the second term is negligible compared with the first. 
Similarly, the normalization factors of the fourth term $\sim 1/L^2$ and the $\sim L$ appearing from the integration result in a $\sim 1/L$ factor, and thus it is negligible. 
The physical interpretation is clear; two electrons in the localized mode have strong repulsion, whereas the modes in the bulk spread in the whole region, and thus the Coulomb repulsion is irrelevant. 

Next, we consider the electrons in the substrate, for which we assume the free electrons 
\begin{align}
H_{\rm substrate}=&\sum_{k, \sigma} \epsilon_k  \hat a_{k,\sigma}^\dagger\hat a_{k,\sigma},   
\label{eq:substrate}
\end{align}
where $\hat a_{k,\sigma}$ ($\hat a_{k,\sigma}^\dagger$) represent the annihilation (creation) operator of the mode with energy $\epsilon_k$ with spin $\sigma$ in the substrate. 
We also consider the tunnel coupling between the system and the substrate as 
\begin{align}
H_{\rm tunnel}=&\int dx \sum_{k, \sigma} V \hat a_{k,\sigma}^\dagger e^{-ikx}\hat \psi_{\sigma}(x)+{\rm h.c.},  
\label{eq:tunnel}
\end{align}
where $V$ is the strength of the tunnel coupling. 
Using the mode-expansion \eqref{eq:modeexpansion} and repeating a similar argument with the interaction term,  we obtain the following effective interaction 
\begin{align}
&H_{\rm tunnel}^{\rm eff}=\sum_{k, \sigma} \left(V_{k}^{{\rm eff}}  \hat a_{k,\sigma}^\dagger \hat d_{\sigma}+{\rm h.c.}\right), \label{eq:tunneleff}\\
&V_{k}^{\rm eff}:=\sqrt{m}\int dx  V  e^{-m|x|-ikx}. \label{def:tunneleff}
\end{align}
This effective tunneling $V_{k}^{\rm eff}$ can be further evaluated as 
\begin{align}
V_{k}^{\rm eff}
&=\frac{2m\sqrt{m}}{m^2+k^2}V. 
\label{eq:tunneleff}
\end{align}

\section{Comparison with a discretized lattice model}
\label{discretemodel}

In the main text we derived the effective Anderson impurity model starting from a continuum Dirac Hamiltonian with a domain-wall mass. 
One may wonder whether the resulting structure, in particular the energy-dependent hybridization and the scaling with the topological mass $m$, is an artifact of the continuum description.

\begin{figure}
\centering
\includegraphics[clip,width=5cm]{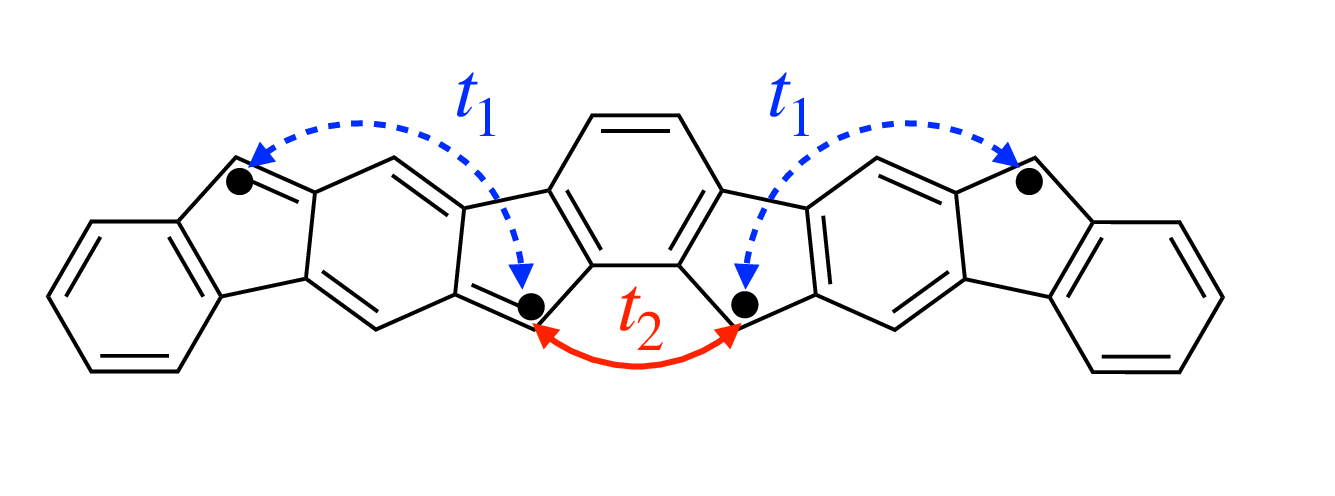}
\caption{A double isomeric Class-II oligo (indenoindene)  as an SSH model with a strong Coulomb interaction. Due to the difference in the distance, $t_2> t_1$ is naturally realized.  
}
\label{Fig:SSHModel}
\end{figure}

To demonstrate that the mechanism is generic, we repeat the construction starting from a lattice model that hosts a topological soliton. 
As a minimal example, we consider the Su-Schrieffer-Heeger (SSH) chain, which is known to support a localized soliton state at a domain wall, with Coulomb repulsion realized, for instance, in a double isomeric Class-II oligo (indenoindene) \cite{SSH,Heeger, Ortiz2025, Aligia2025, Yoshii2026} (See Fig.~\ref{Fig:SSHModel}). 

The SSH Hamiltonian is given by 
\begin{equation}
H_{\mathrm{SSH}} =
\sum_{n,\sigma}
\left(
t_1 \hat c^\dagger_{n,A,\sigma} \hat c_{n,B,\sigma}
+
t_2 \hat c^\dagger_{n,B,\sigma} \hat c_{n+1,A,\sigma}
+ h.c.
\right).
\label{eq:SSH}
\end{equation}
For $|t_2|>|t_1|$ the system supports a topological soliton localized near a domain wall.
The corresponding wavefunction has the form 
\begin{equation}
\psi_{\mathrm{soliton}}(i)
\simeq
\sqrt{\frac{1-r^2}{2}}\, r^{|i-N_s|},
\end{equation}
where $r=\sqrt{|t_1|/|t_2|}$ and $N_s$
denotes the position of the domain wall.

A similar discussion with the present work yields the Anderson model, where the tunneling between the soliton mode in the SSH chain and the metallic substrate becomes 
\begin{equation}
\tilde V_k =
V_0
\sum_n
e^{ik(n-N_s)}
\sqrt{\frac{1-r^2}{2}}
r^{|n-N_s|}.
\end{equation}
Evaluating the geometric series gives
\begin{equation}
\tilde V_k =
V_0
\frac{\sqrt{2(1-r^2)}(1-r^2)}{1+r^2-2r\cos k}.
\end{equation}

Expanding near the continuum limit
$r \rightarrow 1$,
the lattice parameter $r$ is related
to the Dirac mass through
\begin{equation}
\tilde m \propto (1-r^2)\sim 2(1-r). 
\end{equation}
This relation gives the following effective coupling for small $k$ 
\begin{equation}
\tilde V_k \sim 
V_0
\frac{\sqrt{2(1-r^2)}(1-r^2)}{(1-r)^2+rk^2}\propto \frac{\sqrt{2m}m}{m^2+c k^2}, 
\end{equation}
where $c$ is a constant depending on the macroscopic parameters. 
Therefore, the lattice model reproduces the
same filter structure obtained in the
continuum Dirac theory. 
This shows that the topology-controlled
impurity physics discussed in the main text
is not an artifact of the continuum
approximation but a generic consequence of
solitonic defect states in one-dimensional
topological systems.

\section{Evaluation of the logarithmic renormalization integral}
\label{RGevaluation}

In this Appendix, we explicitly evaluate the logarithmic integral that determines the effective ultraviolet scale of the Kondo problem in the presence of the soliton-induced form factor.

The renormalization of the exchange coupling involves the integral
\begin{equation}
I(D) = \int_0^D \frac{d\epsilon}{\epsilon} \mathcal{F}(\epsilon,m),
\end{equation}
where the form factor is given by
\begin{equation}
\mathcal{F}(\epsilon,m) = \frac{1}{\left[1+(\epsilon/m)^2\right]^2}.
\end{equation}
We have factored out the overall scale and focus on the dimensionless form factor. 

We introduce the dimensionless variable $x = \epsilon/m$, which yields
\begin{equation}
I(D) = \int_0^{D/m} \frac{dx}{x} \frac{1}{(1+x^2)^2}.
\end{equation}
This integral can be evaluated analytically. Using partial fraction decomposition, we obtain
\begin{equation}
I(D)
= \ln x - \frac{1}{2}\ln(1+x^2)+\frac{1}{2(1+x^2)} \Bigg|_{x=\Lambda}^{x=D/m},
\end{equation}
where $\Lambda = T_K/m$ serves as the infrared cutoff. 
Taking the limit $D \gg m$ (i.e., $x \to \infty$), we find
\begin{equation}
\ln x - \frac{1}{2}\ln(1+x^2) \to 0,
\quad
\frac{1}{2(1+x^2)} \to 0,
\end{equation}
so that the upper limit gives a finite constant contribution. 

At the lower limit $x = T_K/m \ll 1$, we obtain
\begin{equation}
I(D) = \ln\frac{m}{T_K} + C + \mathcal{O}\left(\frac{T_K^2}{m^2}\right),
\end{equation}
where $C$ is a finite constant independent of $D$. 
The constant $C$ only renormalizes the prefactor of $T_K$ and does not affect the exponential dependence.  
Thus, the logarithmic renormalization takes the form
\begin{equation}
I(D) = \ln\frac{m}{T_K} + \mathcal{O}(1),
\end{equation}
showing that the logarithmic growth is terminated around $m$. 
This shows that the soliton mass acts as an effective ultraviolet scale controlling the logarithmic renormalization. 
Importantly, this cutoff is not sharp: the suppression of high-energy contributions arises from the smooth decay of the form factor $F(\epsilon,m) \sim \epsilon^{-4}$ for $\epsilon \gg m$. 
Nevertheless, the logarithmic contribution is dominated by the energy window $\epsilon \lesssim m$, establishing $m$ as an emergent ultraviolet scale for the Kondo effect.

This demonstrates that the logarithmic renormalization is not sensitive to the bare bandwidth $D$, but instead saturates at the scale $m$. 
The soliton mass, therefore, acts as an emergent ultraviolet scale, arising from the spatial structure of the wavefunction rather than an explicit cutoff.

\section{Result for parabolic dispersion}
\label{parabolic}

In the main text, we assume the linear dispersion for the bulk system (metallic substrate). 
Although the qualitative behavior remains unchanged, we present the result for the case in which the bulk dispersion is parabolic. 
In the case of the parabolic dispersion $\epsilon=B^2k^2$, the topological filter as a function of $\epsilon$ becomes, 
\begin{equation}
\mathcal{F}(\epsilon, m)=\left(\frac{m^2}{m^2+\epsilon/B^2}\right)^2=\frac{1}{\left[1+\epsilon/{\overline m}^2\right]^2}, 
\label{topfilterpara}
\end{equation}
where we define the scaled topological mass as ${\overline m}:=Bm$. 

\begin{figure}[!t]
\centering
\includegraphics[clip,width=9cm]{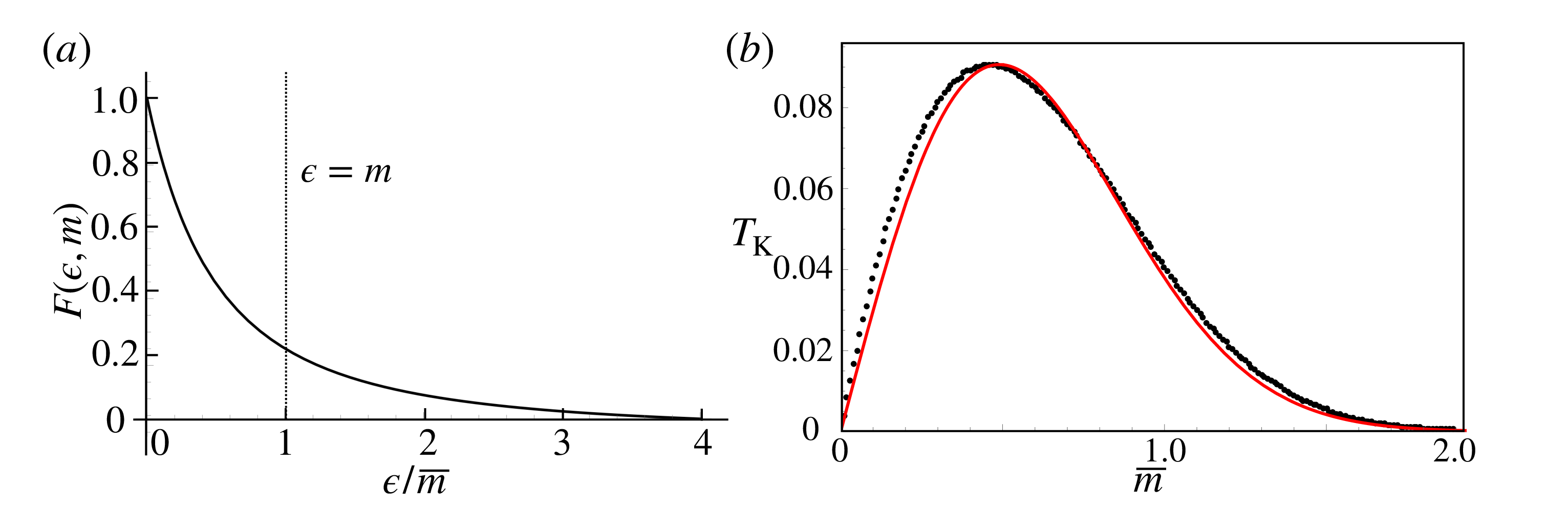}
\caption{Topological filter as a function of $\epsilon/\overline{m}$ (Fig.~(a)) and Kondo temperature $T_{\rm K}$ as a function of the scaled topological mass (Fig.~(b)). 
The black dots represent the numerical scaling flow that starts from the initial coupling $g_0=\rho J(m)$ for $g=2.0$ and $\rho V^2 = 0.15$, where we assume the dispersion relation in the bulk is given as $\epsilon=B^2k^2$ and define the scaled topological mass as ${\overline m}=Bm$. 
The red solid line shows the approximated expression given in Eq.~\eqref{TKapproximated}, and the dimensionless constant is chosen such that the peak value coincides with the numerical result. 
}
\label{fig:TKVSm_parabolic}
\end{figure}

In Fig.~\ref{fig:TKVSm_parabolic}, we show the Kondo temperature obtained for the substrate with the parabolic dispersion relation. 
The black dots represent the numerical scaling flow based on Eq.~\eqref{poormanscaling} with the topological filter given by Eq.~\eqref{topfilterpara} for the parabolic dispersion. 
The solid red line shows the approximated expression given in Eq.~\eqref{TKapproximated}, where we make a replacement $m\to \overline m$.

\section{Comparison with s-d model}
\label{sdmodel}
\begin{figure}[!t]
\centering
\includegraphics[clip,width=5cm]{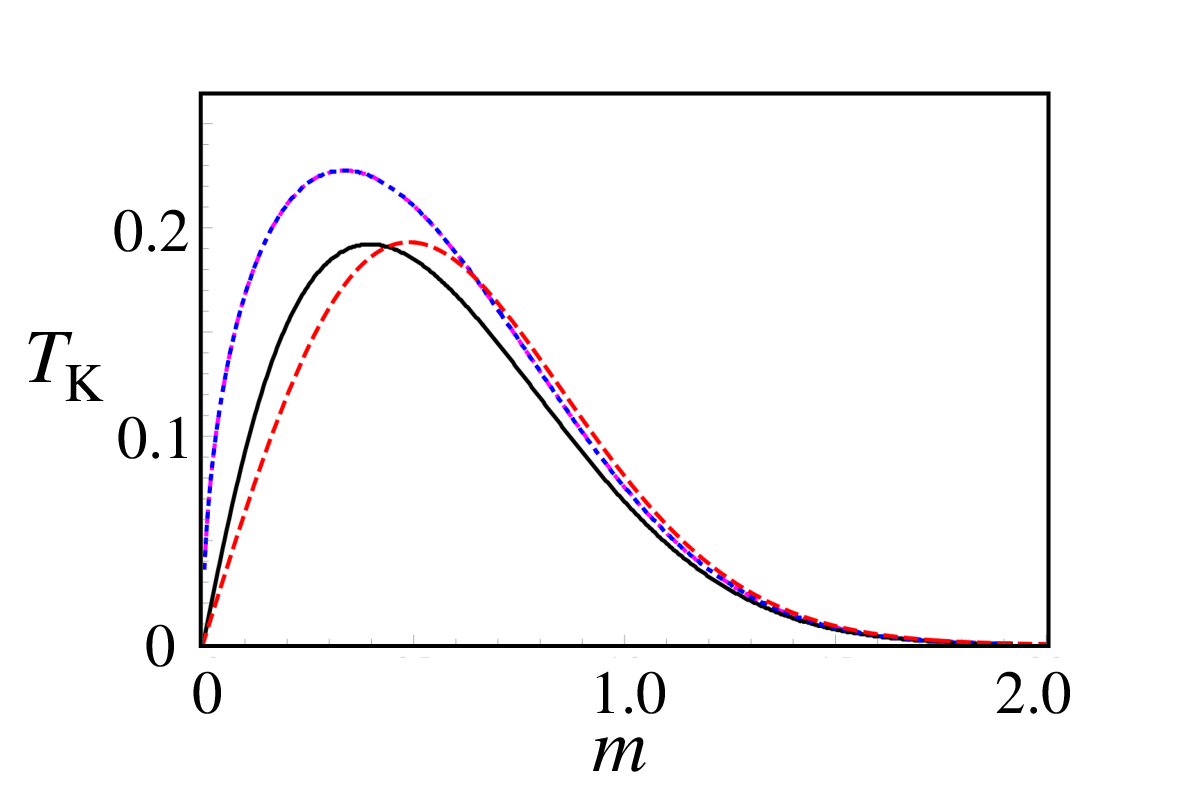}
\caption{Kondo temperature $T_{\rm K}$ as a function of the topological mass $m$ for $g=2.0$ and $\rho V^2 = 0.15$. 
The black dots and red dashed line correspond to the results already shown in Fig.~\ref{fig:TKVSm}. 
The blue dotted and magenta dot-dashed lines, which completely overlap, are obtained by directly integrating Eq.~\eqref{sdmodelscaling} for band widths $D=10$ and $D=20$, respectively. 
}
\label{fig:TKLargeD}
\end{figure}

In the main text, we consider the Anderson model and carry out the two-stage scaling calculation. 
Here we present the consistency with the calculation based on the s-d model. 
Since the localized solitonic state has a finite spatial extent, we need to change the usual delta function type interaction to the following interaction
\begin{align}
&J\sum_{\sigma,\sigma^\prime,\tau,\tau^\prime}\int dx d^\dagger_{\sigma}(\hat {\bf S}_d)_{\sigma,\sigma^\prime} d_{\sigma}  \phi(x)^2 \cdot \hat \chi_{\tau} (x) (\hat {\bf s})_{\tau,\tau^\prime} \hat \chi_{\tau^\prime}(x)\nonumber\\
&=mJ\int dx e^{-2m|x|-ikx+ik^\prime x}\nonumber\\ 
&\quad \times \sum_{k,k^\prime,\sigma,\sigma,\tau,\tau^\prime}d^\dagger_{\sigma}(\hat {\bf S}_d)_{\sigma,\sigma^\prime} d_{\sigma}  \cdot \hat c_{k,\tau}^\dagger (\hat {\bf s})_{\tau,\tau^\prime}\hat c_{k^\prime ,\tau^\prime},  
\end{align}
where $\hat {\bf S}_d$ and $\hat {\bf s}$ represent the vectors whose components are the Pauli matrices.  
The resulting exchange coupling becomes 
\begin{align}
&J_{k,k^\prime}=mJ\int dx e^{-2m|x|-ikx+ik^\prime x}=\frac{m^2}{m^2+q^2} J,  
\end{align}
where we define $q:=(k-k^\prime)/2$ for simplicity. 
The exchange coupling acquires a form factor $J_{kk'} = J f(q)$, reflecting the spatial structure of the soliton wavefunction. 
This leads to a modified scaling equation
\begin{equation}
\frac{dJ}{d\ln E}
= -\rho J^2\, \mathcal{F}\!\left(\frac{E}{m}\right),
\label{sdmodelscaling}
\end{equation}
where $\mathcal{F}(s)$ approaches a constant for $s \ll 1$ and decays rapidly for $s \gg 1$. 
As a result, the
renormalization flow is effectively cut off at the energy scale set by the soliton mass $m$, yielding
$T_K \sim m\, e^{-1/\rho J}$.

Fig.~\ref{fig:TKLargeD} shows the Kondo temperature $T_{\rm K}$ as a function of the topological mass $m$ for $g=2.0$ and $\rho V^2 = 0.15$ obtained by directly integrating Eq.~\eqref{sdmodelscaling} for band widths $D=10$ (blue dotted line) and $D=20$ (magenta dot-dashed line). 
For comparison, we also put two lines (black solid and red dashed lines) already plotted in  Fig.~\ref{fig:TKVSm}. 
All results are semi-quantitatively coincident, and we find that $D=10$ and $D=20$ yield the same results. 
It shows that the topological filter indeed makes a high-energy irrelevant in the present model.


\end{document}